\documentclass[aps,pre,twocolumn,preprintnumbers,amsmath,amssymb,superscriptaddress]{revtex4}
\usepackage{amsmath}
\usepackage{amssymb}

\usepackage{epsfig}
\usepackage{graphicx}
\usepackage{dcolumn}
\usepackage{bm}
\usepackage{color} 
\usepackage{hyperref}

\begin{document}

\title{Some Mathematical Models for ELM Signal}
\author{Hua-sheng XIE} 
\email{huashengxie@gmail.com}
\affiliation{Institute for Fusion
Theory and Simulation, Zhejiang University, Hangzhou 310027,
People's Republic of China}
\date{\today}

\begin{abstract}
There is no wide accepted theory for ELM (Edge Localized Mode) yet.
Some fusion people feel that we may never get a final theory for ELM
and H-mode, since which are too complicated (also related to the
unsolved turbulence problem) and with at least three time scales.
The only way out is using models. (This is analogous to that we
believe quantum mechanics can explain chemistry and biology, but no
one can calculate DNA structure from Schrodinger equation directly.)
This manuscript gives some possible mathematical approaches to it. I
should declare that these are just math toys for me yet. They may
inspire to good understandings of ELM and H-mode, may not. Useful or
useless, I don't know. One need not take too much care of it. Just
for fun and enjoying different interesting ideas.
\end{abstract}
\maketitle

\section{Introduction}
\begin{figure}[htp]
\includegraphics[clip=true,width=8.5cm]{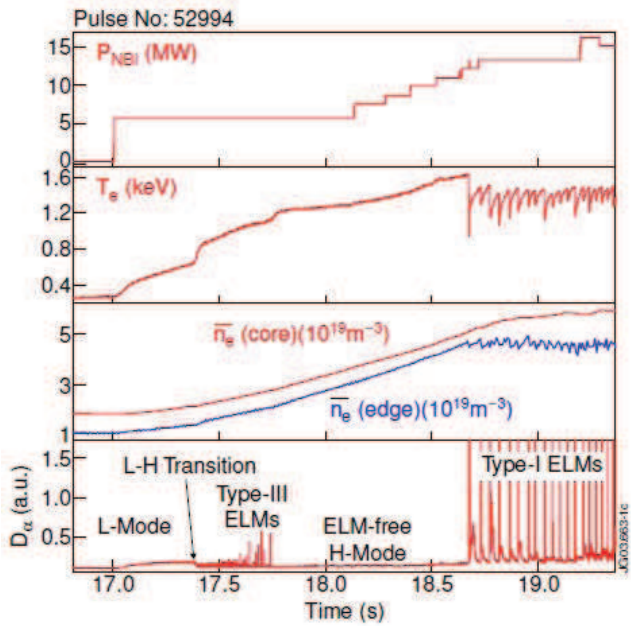}
\caption{\label{fig1} JET typical ELM signals ([Perez2004,
thesis]).}
\end{figure}

Typical ELM signals are shown in FIG.\ref{fig1}. Here, we try to
model type-I and type-III ELM signals of $D_\alpha$.

We just focus on the signal itself and ignore all the other
information behind, i.e., we just discuss how to use (as simple as
possible) equations or other approaches to reproduce the shape of
these signals. It is qualitative or at most semi-quantitative.

\section{Ordinary Differential Equation (ODE)}
This is a wide used approach for modeling. For example, the famous
prey-predator model for fishbone and drift wave-zonal flow system.
One may also extend it to PDE to contain the spatial information.
Indeed, this way has been used to model ELM and L-H transit in
literatures, e.g., [Diamond1994], [Itoh1991, 1993, 1999].

If we assume that the dynamic of signal is not explicit depend on
time $t$, then one equation $\dot{y}=f(y)$ is not enough because
that $\dot{y}$ is not single valued with $y$. So, at least, we must
use second order equation or use two first order equations. While,
one second order equation is just a special case of two first order
equations.

\subsection{Example 1}
In this example, we use the second option, i.e., two first order
equations, which is also used in [Diamond1994]. The equations are
\begin{eqnarray}\label{eq:ode1}
  {{\dot y}_1} = {a_1}{y_1} - {b_1}y_1^2 - {b_2}{y_1}{y_2}, \cr
  {{\dot y}_2} =  - {a_2}{y_2} + {b_3}{y_1}{y_2}.
\end{eqnarray}

\begin{figure}[htp]
\includegraphics[clip=true,width=8.5cm]{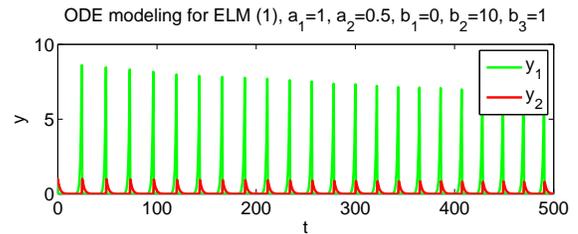}
\caption{\label{fig:ode1} ODE for modeling ELM, example 1.}
\end{figure}

A result is shown in FIG. \ref{fig:ode1}.

The physical explanations of the equations and parameters can be
found in [Diamond1994], however, where $y_1=E$, $y_2=U$.

\subsection{Example 2}
\begin{figure}[htp]
\includegraphics[clip=true,width=8.5cm]{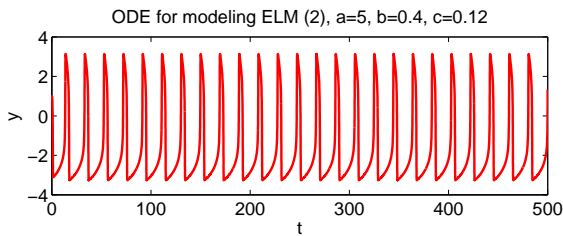}
\caption{\label{fig:ode2} ODE for modeling ELM, example 2.}
\end{figure}

Using one second order equation, e.g.,
\begin{equation}\label{eq:ode2}
\ddot y = a(1 - b{y^2})(\dot y + c) - y - ac,
\end{equation}
which is modified from stiff system equation, a result is shown in
FIG. \ref{fig:ode2}.

The shape is very similar to the PDE result in [Itoh1993]. But, an
apparent drawback of (\ref{eq:ode2}) is that the signal $y$ is not
always positive here.

\section{Delay Differential Equation (DDE)}
DDEs contain derivatives which depend on previous time. We guess
some equations here. The equations are from or modified from
[Shampine2000]. There is also a type of DDE has sawtooth solutions
(see e.g., [Mallet-Paret2011]), which may be also suitable for model
the unsolved sawteeth phenomena in magnetic confined fusion study.

\subsection{Example 1}
The equation is
\begin{eqnarray}\label{eq:dde1}
y'(t) =  - \lambda y(t - 1)( 1 + y( t)), \cr y(t) = t, t \le 0.
\end{eqnarray}

\begin{figure}[htp]
\includegraphics[clip=true,width=8.5cm]{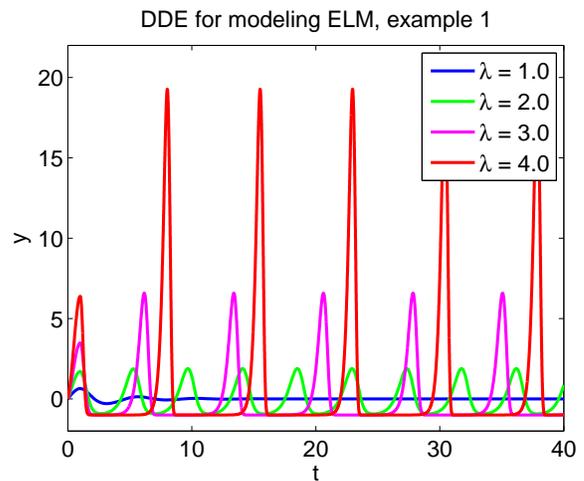}
\caption{\label{fig:dde1} DDE for modeling ELM, example 1.}
\end{figure}

Results are shown in FIG. \ref{fig:dde1}.

\subsection{Example 2}
Equation is
\begin{eqnarray}\label{eq:dde1}
y'(t ) = ry(t)\left( {1 - {{y\left( {t - 0.74} \right)} \over m}}
\right), \cr y(t) = 19, t < 0; y(0) = 19.001.
\end{eqnarray}
which is original to model four-year cycle of the population of
lemmings.

\begin{figure}[htp]
\includegraphics[clip=true,width=8.5cm]{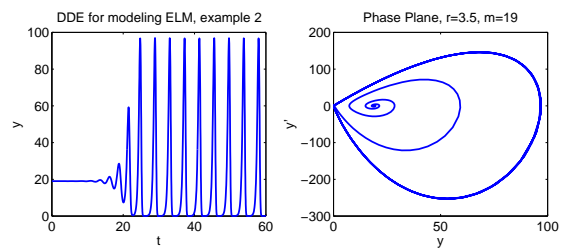}
\caption{\label{fig:dde2} DDE for modeling ELM, example 2.}
\end{figure}

The result is shown in FIG. \ref{fig:dde2}.

\section{Differential Equation + Monte Carlo (DEMC)}
This is a more intuitive way to model ELM, which combines different
time scales using a more acceptable approach. The burst of ELM
signal is suggested related to peeling-ballooning mode in standard
ELM model ([Connor1998]). So, we give two time scale: transport time
scale and ballooning mode time scale.

To simplify the transport process, the transport time scale is
modeled using
\begin{equation}\label{eq:demc1}
{d \over {dt}}f(t) = {f \over {{\tau _T}}}.
\end{equation}

When pressure gradient exceeds marginal value (using $f_{c1}$ to
represent it), the ballooning mode occurs. The signals will grow
exponentially
\begin{equation}\label{eq:demc2}
{d \over {dt}}f(t) = {\gamma _{BM}}f,
\end{equation}
with the characteristic time $\tau_{BM}=1/\gamma_{BM}$.

\begin{figure}[htp]
\includegraphics[clip=true,width=8.5cm]{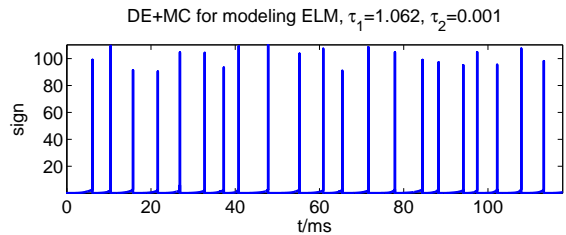}
\caption{\label{fig:demc} DEMC modeling for ELM.}
\end{figure}

When $f(t)$ exceeds a marginal value $f_{c2}$, the plasma crash. And
then, go to next cycle. We also add some randomicities to $f_{c1}$
and $f_{c2}$ to make the whole signal more like the actual
experimental signal. A result is shown in FIG. \ref{fig:demc}.

This simple DEMC model captures some fundamental properties of
integrated simulations, such as [Chang2008] and [Lonnroth2009].

\section{Summary and More}
Which one is more reasonable? At present, I choose DEMC $>$ ODE $>$
DDE. While, maybe none of them can be final explanation of ELM or
H-mode.

There is another math tool called Fractional Differential Equation
(FDE), which has been used for anomalous transport/diffusion
problems in some areas. For plasma community, FDE is not wide
accepted yet, but it is possible in the future. FDE is hard to give
periodic solutions as ELM at present. So, I haven't given examples.


\begin{thebibliography}{99}

\bibitem{Chang2008} Chang, C. S.; Klasky, S.; Cummings, J.; Samtaney, R.;
Shoshani, A.; Sugiyama, L.; Keyes, D.; Ku, S.; Park, G.; Parker, S.;
Podhorszki, N.; Strauss, H.; Abbasi, H.; Adams, M.; Barreto, R.;
Bateman, G.; Bennett, K.; Chen, Y.; Azevedo, E. D.; Docan, C.;
Ethier, S.; Feibush, E.; Greengard, L.; Hahm, T.; Hinton, F.; Jin,
C.; Khan, A.; Kritz, A.; Krsti, P.; Lao, T.; Lee, W.; Lin, Z.;
Lofstead, J.; Mouallem, P.; Nagappan, M.; Pankin, A.; Parashar, M.;
Pindzola, M.; Reinhold, C.; Schultz, D.; Schwan, K.; Silver, D.;
Sim, A.; Stotler, D.; Vouk, M.; Wolf, M.; Weitzner, H.; Worley, P.;
Xiao, Y.; Yoon, E. and Zorin, D., Toward a first-principles integrated
simulation of tokamak edge plasmas, Journal of Physics: Conference
Series, 2008, 125, 012042.

\bibitem{Connor1998} Connor, J. W.; Hastie, R.
J.; Wilson, H. R. and Miller, R. L., Magnetohydrodynamic stability of
tokamak edge plasmas, Physics of Plasmas, 1998, 5, 2687-2700.


\bibitem{Diamond1994} Diamond, P. H.; Liang, Y.-M.; Carreras, B. A. and Terry,
P. W., Self-Regulating Shear Flow Turbulence: A Paradigm for the L
to H Transition, Phys. Rev. Lett., 1994, 72, 2565-2568.

\bibitem{Itoh1991}
Itoh, S.-I.; Itoh, K.; Fukuyama, A. and Miura, Y., Edge localized mode
activity as a limit cycle in tokamak plasmas, Phys. Rev. Lett.,
1991, 67, 2485-2488.

\bibitem{Itoh1993} Itoh, S.-I.; Itoh, K. and Fukuyama,
A., The ELMy H mode as a limit cycle and the transient responses of
H modes in tokamaks, Nuclear Fusion, 1993, 33, 1445.

\bibitem{Itoh1999}
Itoh, K.; Itoh, S. I. and Fukuyama, A., Transport and Structural
Formation in Plasmas, IOP, 1999.

\bibitem{Lonnroth2009} Lonnroth, J.-S., A
Predictive Modelling of Edge Transport Phenomena in ELMy H-Mode
Tokamak Fusion Plasmas, Helsinki University of Technology, PhD
thesis, 2009.

\bibitem{Mallet-Paret2011} Mallet-Paret, J. and Nussbaum, R. D.,
Superstability and rigorous asymptotics in singularly perturbed
state-dependent delay-differential equations, Journal of
Differential Equations, 2011, 250, 4037 - 4084.

\bibitem{Perez2004} Perez,
C., MHD analysis of edge instabilities in the JET tokamak,
University of Utrecht, PhD thesis, 2004.

\bibitem{Shampine2000} L. F.
Shampine, Solving Delay Differential Equations with dde23,
Mathematics Department Southern Methodist University, 2000.


\end{thebibliography}

\end{document}